\documentclass[letterpaper,prb,twocolumn,nobibnotes]{revtex4}
\usepackage{epsfig}
\usepackage{amsmath}
\usepackage{fancybox,amssymb}
\renewcommand{\eqref}[1]{(\ref{#1})}

\begin{document}
\title{\bf Mapping of Dissipative Particle Dynamics in Fluctuating Hydrodynamics Simulations}
\author{R. Qiao}
 \email[Corresponding author. Email: ] {rqiao@ces.clemson.edu} 
% \homepage{http://www.clemson.edu/~rqiao}
\author{P. He}
\affiliation{ %Department of Mechanical Engineering, 
             College of Engineering \& Sciences, Clemson University, Clemson, SC, 29634}
\date{\today}
\begin{abstract}
Dissipative particle dynamics (DPD) is a novel particle method for
mesoscale modeling of complex fluids. DPD particles are often thought
to represent packets of real atoms, and the physical scale probed in
DPD models are determined by the mapping of DPD variables to the
corresponding physical quantities. However, the non-uniqueness of
such mapping has led to difficulties in setting up simulations to
mimic real systems and in interpreting results. For modeling transport
phenomena where thermal fluctuations are important (e.g., fluctuating
hydrodynamics), an area particularly suited for DPD method, we propose
that DPD fluid particles should be viewed as only 1) to provide a
medium in which the momentum and energy are transferred according to
the hydrodynamic laws and 2) to provide objects immersed in the DPD
fluids the proper random "kicks" such that these objects exhibit
correct fluctuation behaviors at the macroscopic scale. We show that,
in such a case, the choice of system temperature and mapping of DPD 
scales to physical scales are uniquely determined by the level of 
coarse-graining and properties of DPD fluids. We also verified that 
DPD simulation can reproduce the macroscopic effects of thermal 
fluctuation in particulate suspension by showing that the Brownian 
diffusion of solid particles can be computed in DPD simulations 
with good accuracy. 

\end{abstract}
\maketitle

%% ==============
%% Introductions
%% ==============
%\section{Introduction}
Dissipative particle dynamics (DPD) is a method developed primarily for
the simulation of complex fluids at mesoscopic scales \cite{hoo,espanol1,groot}. 
While it is often thought that DPD beads represent packets of real atoms moving 
collectively, the statistical mechanical foundation of such a view for DPD 
model remains obscure \cite{pre}. Associated with the ambiguity of the exact 
nature of DPD beads is the ambiguity in mapping of DPD scales to the physical 
scales. While the mapping of length is straightforward, the mapping of time 
is more complicated. Depending on the problems being studied, mapping of DPD 
time to physical time has been based on bead diffusion rate, 
bead thermal velocity, or externally imposed time scales \cite{diff,george,pump}, 
to name just a few. In all but a few cases \cite{pump}, temperature of the 
DPD system was chosen arbitrarily. Established procedures for mapping 
between DPD and physical scales and for choosing system temperature are not 
yet available.

Our interest in DPD originates from the need to study transport phenomena
in particulate suspension where thermal fluctuations may play a critical 
role \cite{phelan,molSimu}. Such transport can be described by the fluctuating 
hydrodynamics theories \cite{fhd}. However, solving the hydrodynamics equations 
in particulate suspension is computationally demanding and introducing thermal 
fluctuations that rigorously satisfy the Fluctuation-Dissipation Theorem (FDT) 
is challenging \cite{Patankar}. DPD is well-suited for studying such phenomena 
as momentum/energy conservation and FDT are guaranteed by the way DPD models are 
designed, and colloidal particles can be modeled easily by bonding DPD beads 
together \cite{espanol1,groot}. However, there are two unresolved issues 
in DPD simulation of fluctuating hydrodynamics, namely, how to set up a model 
for a given physical system and how to interpret the simulation results. For 
example, to study the diffusion of a 30 nm diameter particle in water at 300 K, 
what should be the temperature of the DPD system and how to map the results 
to dimensional values are not clear.
% Addressing this question is the focus of this paper. 

%The rest of the paper is organized as follows: Section \ref{sec:map} discusses 
%the proposed mapping of DPD model and the implication on the choice of system 
%temperature, Section \ref{sec:example} applies the proposed method for mapping 
%and selecting system temperature to the study of Brownian motion of a colloidal 
%particle. Finally, conclusions are presented in Section \ref{sec:conclusions}.

%\section{Mapping of DPD model and choice of system temperature}
%\label{sec:map}

To address the above issues, we propose to abandon the idea that DPD fluid 
beads are ``clumps'' of real fluid atoms, but view them \textit{together}
as a ``media'' that provides an arena for the transport of momentum, energy 
and particulates that satisfies the fluctuating hydrodynamics laws. This idea 
is inspired by a recent paper on coarse-graining in colloidal suspensions \cite{pre}. 
To bridge the DPD and physical systems, we require that
\begin{enumerate}
\item The time scale of diffusional transport of momentum 
  (and energy if in non-isothermal simulations) inside the DPD fluids 
  should match that of the real fluids;
\item DPD fluids should provide objects immersed in them the proper random 
  "kicks" such that these objects exhibit correct fluctuation behaviors at the 
  macroscopic scale.
\end{enumerate}

We now consider the application of the above requirements in the simulation of 
colloidal particles (diameter: $\overline{d}$) dispersed in 
fluids (density: $\overline{\rho}$, kinematic viscosity: $\overline{\nu}$, 
temperature: $\overline{T}$). Properties of 
corresponding DPD fluids are denoted by the same symbol as in real fluids but 
without the bar, e.g., density of DPD fluids is denoted as $\rho$. 
We will limit our discussion to isothermal simulations, and the extension
to non-isothermal simulations is straightforward. The DPD model reads \cite{groot}
%\begin{eqnarray}
\begin{equation}
    d \boldsymbol{r}_i  =  \boldsymbol{v}_i dt; \ \ \ 
  m d \boldsymbol{v}_i  =  \boldsymbol{F}^C_i dt + \boldsymbol{F}^{D}_i dt + \boldsymbol{F}^{R}_i \sqrt{dt} 
%\end{eqnarray}
\end{equation}
where $m$, $\boldsymbol{r}_i$, and $\boldsymbol{v}_i$ are the mass,
position, and velocity of bead $i$, respectively. $T$ is the system
temperature. 
$\mathbf{F}^C_i$, $\mathbf{F}^D_i$ and $\mathbf{F}^R_i$ are the 
conservative, dissipative and random forces acting on bead
$i$, respectively. These forces are given by \cite{groot,singapore}
\begin{eqnarray} 
\boldsymbol{F}^C_i &
   = & \sum_{j \neq i} a_{ij} w(r_{ij}/r_c) \boldsymbol{e}_{ij} \label{cons} \\
\boldsymbol{F}^{D}_i &
   = & \sum_{j \neq i} - \gamma_{ij} w_d^2(r_{ij}/r_c^d) (\boldsymbol{e}_{ij} \cdot \boldsymbol{v}_{ij}) \boldsymbol{e}_{ij} \label{disp} \\ 
\boldsymbol{F}^{R}_i &
   = & \sum_{j \neq i} \sigma_{ij} w_d(r_{ij}/r_c^d) \theta_{ij} \boldsymbol{e}_{ij} \label{rand}
\end{eqnarray}
where $a_{ij}$ is the conservative force coefficient and 
$r_{ij} = |\mathbf{r}_{ij}| = | \mathbf{r}_i - \mathbf{r}_j|$.
$w$ and $w_d$ are weighting functions with cutoff distances
of $r_c$ and $r_c^d$, respectively. 
$\boldsymbol{e}_{ij} = \mathbf{r}_{ij}/r_{ij}$, and 
$\boldsymbol{v}_{ij} = \mathbf{v}_i - \mathbf{v}_j$.
$\theta_{ij}$ is a symmetric random variable with zero mean and unit variance.
%Variables $\gamma_{ij}$ and $\sigma_{ij}$ determine the strength 
$\gamma_{ij}$ and $\sigma_{ij}$ 
%determine the strength 
%of the dissipative and random force and they 
are lated by the Fluctuation-Dissipation Theorem as 
$ \gamma_{ij}  =  \sigma_{ij}^2 / 2 k_B T \label{fdt1}$,
where $k_B$ is the Boltzmann constant. 
%Usually $\sigma_{ij}$ is taken as a constant. 
Mass, length, and time in the above model are measured by 
$m$, $r_c$, and $\sqrt{k_B T/m}$, respectively. We assume that the unit length 
and time in DPD map to physical length [L] and time [t], respectively.
%The colloidal particle diameter in DPD simulation is thus $\overline{d} = d/[L]$. 
To satisfy the proposed requirements, we enforce
\begin{eqnarray}
%\frac{\overline{d}^2}{\overline{\nu}} [t]  & = & \frac{d^2}{\nu} \label{momentum} \\
{d^2}/{\nu} [t] & = & {\overline{d}^2}/{\overline{\nu}}  \label{momentum} \\
%\sqrt{\frac{\overline{k_{B}T}}{\overline{\rho} \overline{d}^3}} \frac{[L]}{[t]} & = &\sqrt{\frac{k_B T}{\rho d^3}}\label{fluctuation}
\sqrt{{k_B T}/{\rho d^3}} {[L]}/{[t]} & = &\sqrt{\overline{k_{B}T}/{\overline{\rho} \overline{d}^3}} \label{fluctuation}
\end{eqnarray}
Combining Equs. (\ref{momentum}-\ref{fluctuation})
\begin{eqnarray}
{[t]}     & = & [L]^{2} \nu/\overline{\nu} \label{time}\\
%k_B T     & = &  \frac{\rho \nu^2 }{\overline{\rho} \overline{\nu}^2 }  \frac{\overline{k_BT}}{[L]} \label{temp}
k_B T     & = &  \frac{\rho \nu^2 }{\overline{\rho} \overline{\nu}^2 }  \frac{\overline{k_BT}}{[L]} \label{temp}
\end{eqnarray}
Equs. (\ref{time}-\ref{temp}) can be used to setup and analyze DPD simulations of
fluctuating hydrodynamics. Equ. \eqref{time} provides the mapping of 
time scale in DPD model. Equ. \eqref{temp} indicates that the temperature 
in DPD simulation of fluctuating hydrodynamics 
%cannot be chosen arbitrarily but 
is determined by the level of coarse-graining (represented by [L] and $\rho$) and 
transport properties of DPD fluids (represented by $\nu$). If $\nu$ and $\rho$ are 
known, then the system temperature is uniquely determined. In practice, since 
$\nu$ of DPD fluids is a function of their temperature, 
the temperature of DPD system can only be determined after the  
dependence of $\nu$ on temperature is known. While Equs. \eqref{momentum} 
and \eqref{fluctuation} have been used to map DPD models to physical scales, 
enforcing them \textit{simultaneously} in fluctuating 
hydrodynamics simulations and thus leading to a unique choice of system 
temperature has not been proposed. 

\begin{figure}[hptb]
  \begin{center}
    \epsfig{file=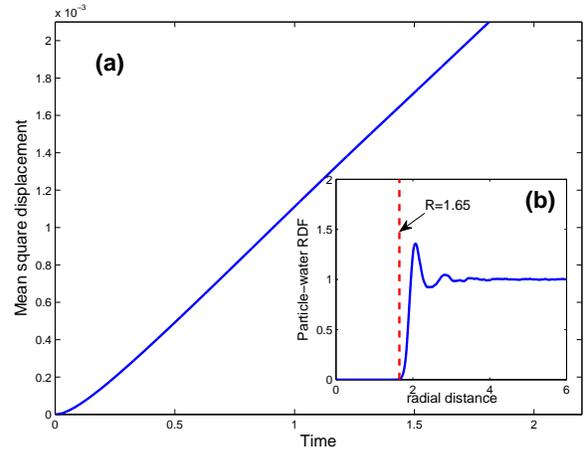,width=3.0in}  
    \caption{(a). Mean square displacement of the nanoparticle immersed in fluids, 
      (b) Particle-fluid pair correlation obtained from DPD simulation. }
    \label{msd}
  \end{center}
\end{figure}

%An important question on the proposed mapping scheme is whether Equs. \eqref{fluctuation} 
%is sufficient to ensure the second requirements mentioned above to be satisfied in
%a DPD simulation. Below we demonstrate that this indeed is the case at least
%under simple conditions. 
To demonstrate the above mapping scheme and to investigate the ability of DPD 
in modeling fluctuating hydrodynamics, we study the diffusion
of a single nanoparticle immersed in fluids with $\bar{\rho} = 10^3$ kg/m$^3$ and 
$\bar{\nu}$ = 0.89$\times$10$^{-6}$ m$^2$/s at 300 K. We set [L]=10 nm and $\rho =$ 6.0. 
The nanoparticle is built by bonding 117 DPD beads together and is modeled as a 
rigid body. In Equ. \eqref{cons}, $w(r_{ij}/r_c) = 1-r_{ij}/r_c$ with 
$r_c =$ 1.0 and $a_{ij}$ is set to $a_{ff}$ = 10.0 and $a_{fp}$ = 17.0, where %subscripts 
$ff$ and $fp$ denote fluid-fluid and fluid-nanoparticle interactions, respectively. 
In Equ. \eqref{disp}, $w_d(r_{ij}/r_c^d) = \sqrt[4]{1-r_{ij}/r_c^d}$ suggested in Ref. \onlinecite{singapore} 
is used with $r_c^d =$ 1.0. $\sigma_{ij}$ in Equ. \eqref{rand} is taken as 5.0 for all
bead pairs. Using these parameters, we first measured the viscosity of 
DPD fluids as a function of temperature, and the temperature in DPD model of 
the nanoparticle-fluids system is then determined via Equ. \eqref{temp} to be 0.3875. 
To investigate the ability of DPD model in reproducing the macroscopic effects of 
thermal fluctuations, we computed the diffusion coefficient ($D_{p}$) of the nanoparticle, 
which is a macroscopic manifest of the thermal fluctuations. Fig. \ref{msd}(a) shows 
the mean square displacement of the particle, and a $D_{p}$ of (1.59$\pm$0.12)$\times$10$^{-4}$ 
was obtained. We also computed $D_{p}$ independently by using the Einstein-Stokes law 
$D_p = k_B T/6 \pi R \mu$, where $\mu$ is the fluid viscosity. From the particle-fluids 
pair correlation function shown in Fig. \ref{msd}(b), the nanoparticle radius $R$ was determined 
to be 1.65, with which a $D_p$ of 1.55$\times$10$^{-4}$ was then computed. Given the ambiguity 
in the definition of particle diameter and the statistical uncertainty of DPD results, 
the agreement between DPD simulation and Einstein-Stokes prediction is very good. This 
verifies the ability of DPD in capturing the macroscopic effects of thermal fluctuations 
in particulate suspensions, which has been assumed but not explicitly confirmed in the literature.

In summary, we proposed a method of mapping DPD simulation of fluctuating hydrodynamics,
and the temperature in such simulations is determined uniquely by the level of coarse-graining 
and properties of DPD fluids. Following the proposed method of choosing system temperature, 
we studied the Brownian diffusion of a nanoparticle and showed that DPD simulation can 
reproduce the macroscopic fluctuation of nanoparticle immersed in fluids with good accuracy.

\bibliography{sources}

\newpage

\end{document}